# Flat FLRW Universe in logarithmic symmetric teleparallel gravity with observational constraints

M. Koussour,[1, *] S. H. Shekh,[2, †] A. Hanin,[1, ‡] Z. Sakhi,[1, 3, §] S. R. Bhoyer,[4, ¶] and M. Bennai[1, 3, **]

[1]*Quantum Physics and Magnetism Team, LPMC, Faculty of Science Ben M'sik,*
*Casablanca Hassan II University, Morocco.*
[2]*Department of Mathematics, S.P.M. Science and Gilani Arts,*
*Commerce College, Ghatanji, Dist. Yavatmal, Maharashtra 445301, India*
[3]*Lab of High Energy Physics, Modeling and Simulations, Faculty of Science,*
*University Mohammed V-Agdal, Rabat, Morocco.*
[4]*Department of Mathematics, Phulsing Naik Mahavidyalaya,*
*Pusad, Yavatmal, Maharashtra-445216, India.*

(Dated: August 26, 2022)

In this paper, we investigate the homogeneous and isotropic flat FLRW Universe in the logarithmic form of $f(Q)$ gravity, where $Q$ is the non-metricity scalar, specifically, $f(Q) = \alpha + \beta \log(Q)$, where $\alpha$ and $\beta$ are free model parameters. In this study, we consider a parametrization of the Hubble parameter as $H(z) = \eta \left[ (z+1)^{-\gamma} + 1 \right]$, where $\gamma$ and $\eta$ are model/free parameters which are constrained by an $R^2$-test from 57 points of the Hubble datasets in the redshift range $0.07 < z < 2.36$. Further, we investigate the physical properties of the model. We analyze the energy conditions to check the compatibility of the model. We found the SEC is violated for the logarithmic form of $f(Q)$ gravity due to the reality that the Universe in an accelerating phase. Finally, we discuss some important cosmological parameters in this context to compare our model with dark energy models such as jerk parameter and statefinder parameters.

## I. INTRODUCTION

At the end of the 1990s, two teams of astronomers began to study the Universe at very great distances, by observing and analyzing the Type Ia supernovae (SN Ia), located at very great distances from us. They called the two High-Z Supernova Search Team (HZSST) and Supernova Cosmology Project (SCP) [1, 2]. Both teams expected the Universe to be in a phase of decelerated expansion, which was confirmed by Friedmann's equations in General Relativity (GR). But the amazing result is that the present Universe is in an accelerated expansion. The simplest explanation for this discrepancy is the cosmological constant ($\Lambda$), which Einstein previously introduced into his field equations to stabilize the Universe. The cosmological constant is an exotic form of energy, they call it Dark Energy (DE). The idea closest to physicists was that this energy was vacuum energy, but problems have arisen on this topic. According to the $\Lambda CDM$ ($\Lambda$ Cold Dark Matter) model and observational data, the proportion of DE in the Universe is 72% of the total density of the Universe, about 23% of dark matter (DM), and only about 5% of baryonic matter. There is an important parameter in cosmology that helps us study DE models, called the equation of state (EoS) parameter $\omega = \frac{p}{\rho}$, where $\rho$ is the energy density of the Universe and $p$ is the pressure. In the $\Lambda CDM$ model $\omega = -1$, meaning the Universe has negative pressure. The new 2018 Planck "TT, TE, EE + lowE + lensing" (where TT, TE and EE respectively refer for temperature angular power spectrum, temperature-polarization cross-power spectrum and cross-frequency power spectrum) measurement gave constraints on the DE EoS parameter as follows [3]:

- $\omega = -1.56^{+0.60}_{-0.84}$ (Planck + TT + lowE),

- $\omega = -1.58^{+0.52}_{-0.41}$ (Planck + TT, EE + lowE),

- $\omega = -1.57^{+0.50}_{-0.40}$ (Planck + TT, TE, EE + lowE + lensing),

- $\omega = -1.04^{+0.10}_{-0.10}$ (Planck + TT, TE, EE + lowE + lensing + BAO).

Despite all these efforts to explain the DE and cosmic acceleration, the mystery of DE remains one of the greatest mysteries of unresolved physics to date. Recently, other interesting alternatives called modified theories of gravity (MTG) have emerged. The idea of these theories is that cosmic acceleration can be the result of a modification in the field equations of GR, and there is no DE. The principle of these theories is to replace the Ricci scalar $R$ in the Einstein-Hilbert action $S_{EH}$ with a generalized function of $R$ or other physical quantity such as $f(R)$ gravity, $f(G)$ gravity (where $G$ is the Gauss-Bonnet invariant), $f(R,T)$ gravity (here $T$ is the trace of the energy-momentum tensor), etc [4–9]. Recently, several geometrical formulations equivalent of GR have appeared, the most famous of which is teleparallel gravity, in which gravitational interactions are described by the concept of torsion $T$, which replaces the concept of curvature in Weitzenbock's space-time. Since the Riemann curvature tensor in Weitzenbock space-time is zero, the overall geometry is flat. The basic idea of this theory is to replace the metric tensor $g_{\mu\nu}$ of the space-time with a set of tetrads vector $e^i_\mu$ and thus known as the teleparallel equivalent of GR (TEGR), or $f(T)$ gravity (here $T$ is the torsion) [10–14]. However, symmetric teleparallel gravity is another modified theory of gravity equivalent to GR suggested by Jimenez et al. recently [15], and it has obtained much attention among cosmologists. In this theory, the geometric framework is of the type of Weyl geometry which is a generalization of Riemannian geometry, in which gravitational interactions are described by a geometrical variable called a non-metricity scalar $Q$ [16]. The non-metricity tensor $Q_{\gamma\mu\nu}$ mathematically describes a variation in vector length in parallel transport, which results in the covariant derivative of the metric tensor being non-zero $Q_{\gamma\mu\nu} = \nabla_\gamma g_{\mu\nu} \neq 0$, it is consequently a tensor field of order three, unlike Riemannian geometry in which only a variation of direction takes place. This concept disappears in the case of Riemannian geometry and can be exploited to research non-Riemannian space-time. There is another important extension of Weyl geometry called Weyl-Cartan geometry where the concept of torsion is introduced.

[*] pr.mouhssine@gmail.com
[†] daˑsalim@rediff.com
[‡] abdeljalil.hanin.2019@gmail.com
[§] zb.sakhi@gmail.com
[¶] sbhoyar68@yahoo.com
[**] mdbennai@yahoo.fr

Generally, in Weyl-Cartan geometry one can present GR through three equivalent geometrical representations: first, using the concept of curvature with zero torsion and non-metricity, second, using the concept of torsion with zero curvature and non-metricity, and finally, using the concept of non-metricity with zero curvature and torsion. Although the $f(Q)$ gravity theory has only been proposed in recent years, it already offers interesting and attractive results in the literature. Koussour et al. studied the cosmic acceleration and energy conditions in $f(Q)$ using the hybrid expansion law (HEL) [17], while Mandal et al. examined the energy conditions to verify the stability of their supposed cosmological models and constraints the model parameters with the current values of cosmological parameters in $f(Q)$ gravity [18]. Shekh discusses the idea of holographic dark energy (HDE) in $f(Q)$ gravity [19]. Furthermore, several works have been discussed in the framework of $f(Q)$ gravity, see [20–25].

This manuscript is organized as follows: In Sec. II, we present the mathematical formalism of $f(Q)$ gravity and then apply it to the FLRW Universe to find the modified Friedman equations. In Sec. III, we consider a parametrization of the Hubble parameter (or deceleration parameter) as $H(z) = \eta \left[(z+1)^{-\gamma} + 1\right]$, where $\gamma$ and $\eta$ are model parameters constrained by an $R^2$-test from 57 data points of the Hubble data set in the redshift range $0.07 < z < 2.36$, to obtain the exact solutions of the field equations in the logarithmic form of $f(Q)$ gravity and obtain the expressions of pressure, density and EoS parameter in terms of redshift. Also, we discussed energy conditions in the same section. In Sec. IV, we present some physical properties of the parametrization of the Hubble parameter. Finally, we discuss our results in Sec. V.

## II. FLRW UNIVERSE IN $f(Q)$ GRAVITY

In the context of differential geometry, the symmetric metric tensor $g_{\mu\nu}$ is used to defines the length of a vector and an asymmetric connection $Y^\gamma{}_{\mu\nu}$ is used to define the covariant derivatives and parallel transport. Hence, in the case of the Weyl-Cartan geometry, the general affine connection can be decomposed into the following three independent components as [16]

$$Y^\gamma{}_{\mu\nu} = \Gamma^\gamma{}_{\mu\nu} + C^\gamma{}_{\mu\nu} + L^\gamma{}_{\mu\nu}, \qquad (1)$$

with

$$\Gamma^\gamma{}_{\mu\nu} \equiv \frac{1}{2}g^{\gamma\sigma}\left(\partial_\mu g_{\sigma\nu} + \partial_\nu g_{\sigma\mu} - \partial_\sigma g_{\mu\nu}\right), \qquad (2)$$

$$C^\gamma{}_{\mu\nu} \equiv \frac{1}{2}T^\gamma{}_{\mu\nu} + T_{(\mu}{}^\gamma{}_{\nu)}, \qquad (3)$$

being the standard definition of the Levi-Civita connection of the metric $g_{\mu\nu}$ and the contorsion tensor, respectively, with the torsion tensor defined as $T^\gamma{}_{\mu\nu} \equiv 2Y^\gamma{}_{[\mu\nu]}$. Further, the disformation tensor $L^\gamma{}_{\mu\nu}$ is derived from the non-metricity tensor $Q_{\gamma\mu\nu}$ as

$$L^\gamma{}_{\mu\nu} \equiv \frac{1}{2}g^{\gamma\sigma}\left(Q_{\nu\mu\sigma} + Q_{\mu\nu\sigma} - Q_{\gamma\mu\nu}\right). \qquad (4)$$

By definition, the non-metricity tensor $Q_{\gamma\mu\nu}$ is (minus) a covariant derivative of the metric tensor with regard to the asymmetric connection $Y^\gamma{}_{\mu\nu}$, i.e. $Q_{\gamma\mu\nu} =$

$\nabla_\gamma g_{\mu\nu}$, and it can be obtained

$$Q_{\gamma\mu\nu} = -\partial_\gamma g_{\mu\nu} + g_{\nu\sigma} Y^\sigma{}_{\mu\gamma} + g_{\sigma\mu} Y^\sigma{}_{\nu\gamma}. \quad (5)$$

As we described in the introduction, the geometrical framework that we use in this present background is to consider flat space-time with zero torsion, which must correspond to a pure coordinate transformation from the trivial connection as shown in [15]. More clearly, the connection can be parameterised as

$$Y^\gamma{}_{\mu\beta} = \frac{\partial x^\gamma}{\partial \xi^\rho} \partial_\mu \partial_\beta \xi^\rho, \quad (6)$$

It is good to mention in Eq. (6) that $\xi^\gamma = \xi^\gamma(x^\mu)$ is an invertible relation and $\frac{\partial x^\gamma}{\partial \xi^\rho}$ is the inverse of the corresponding Jacobian [27]. Thus, it is always possible to get a coordinate system so that the connection $Y^\gamma{}_{\mu\nu}$ vanish. This condition is called coincident gauge and the covariant derivative $\nabla_\gamma$ reduces to the partial derivative $\partial_\gamma$. Hence, in the coincident gauge coordinate, can be obtained

$$Q_{\gamma\mu\nu} = -\partial_\gamma g_{\mu\nu}. \quad (7)$$

The symmetric teleparallel gravity equivalent to GR (STEGR) within coincident gauge coordinates in which $Y^\gamma{}_{\mu\nu} = 0$ and $C^\gamma{}_{\mu\nu} = 0$, and consequently from Eq. (1) the Levi-Civita connection expression can be given in terms of the disformation tensor as $\Gamma^\gamma{}_{\mu\nu} = -L^\gamma{}_{\mu\nu}$. The action in symmetric teleparallel gravity is described as [15]

$$S = \int \sqrt{-g} d^4x \left[\frac{1}{2} f(Q) + \mathcal{L}_m\right], \quad (8)$$

where $f(Q)$ is an arbitrary function of the non-metricity $Q$, $g$ is the determinant of the metric tensor $g_{\mu\nu}$ and $\mathcal{L}_m$ is the matter Lagrangian density. The above action is different from the action in the original paper [15] because the Lagrange multipliers are enforcing the constraints on the curvature and torsion contributions in the action to not appear in the equations of motion indicating that the constrained solutions are a subset of the unconstrained. Hence, the constraints can be imposed by hand after solving for the stationary points of the unconstrained action. In addition, from the above action, GR can be reproduced for the choice of function in the form $f(Q) = -Q$, i.e. for this choice, we regain the so-called STEGR [26]. The non-metricity tensor defined as

$$Q_{\gamma\mu\nu} = \nabla_\gamma g_{\mu\nu} \neq 0, \quad (9)$$

and its two traces are such that

$$Q_\gamma = Q_\gamma{}^\mu{}_\mu, \quad \tilde{Q}_\gamma = Q^\mu{}_{\gamma\mu}. \quad (10)$$

The variation of the action ($S$) in Eq. (8) with respect to the metric, we get the gravitational field equations given by

$$\frac{2}{\sqrt{-g}} \nabla_\gamma(\sqrt{-g} f_Q P^\gamma{}_{\mu\nu}) + \frac{1}{2} f g_{\mu\nu} + f_Q (P_{\nu\rho\sigma} Q_\mu{}^{\rho\sigma} - 2 P_{\rho\sigma\mu} Q^{\rho\sigma}{}_\nu) = -T_{\mu\nu}. \quad (11)$$

where $f_Q = \frac{df}{dQ}$, the superpotential tensor (or non-metricity conjugate) $P^\gamma{}_{\mu\nu}$ in terms of non-metricity tensor is written as $4P^\gamma{}_{\mu\nu} = -Q^\gamma{}_{\mu\nu} + Q_\mu{}^\gamma{}_\nu + Q_\nu{}^\gamma{}_\mu + (Q^\gamma - \tilde{Q}^\gamma) g_{\mu\nu} - \frac{1}{2}(\delta^\gamma_\mu Q_\nu + \delta^\gamma_\nu Q_\mu)$, and the trace of non-metricity tensor has the form $Q = -Q_{\gamma\mu\nu} P^{\gamma\mu\nu}$. In Eq. (11), $T_{\mu\nu}$ constitute the energy-momentum tensor to the matter Lagrangian is given by

$$T_{\mu\nu} = (\rho + p) u_\mu u_\nu + p g_{\mu\nu}, \quad (12)$$

where $\rho$ and $p$ are the energy density and pressure of the matter content. In addition, we can also take the variation of (8) with respect to the connection, which gives

$$\nabla_\mu \nabla_\gamma \left(\sqrt{-g} f_Q P^\gamma{}_{\mu\nu}\right) = 0. \quad (13)$$

We consider the homogeneous, isotropic and flat Friedmann-Lemaître-Robertson-Walker (FLRW) metric in the form

$$ds^2 = -dt^2 + a^2(t)(dx^2 + dy^2 + dz^2), \quad (14)$$

where $a(t)$ is the scale factor of the model. The field equations (11) and Eq. (12) for the metric (14) are obtained as

$$3H^2 = \frac{1}{2f_Q}\left(-\rho + \frac{f}{2}\right), \quad (15)$$

$$\dot{H} + 3H^2 + \frac{\dot{f}_Q}{f_Q} H = \frac{1}{2f_Q}\left(p + \frac{f}{2}\right), \quad (16)$$

where $H$ is the Hubble parameter defined by $H = \frac{\dot{a}}{a}$, the point designates the derivative with respect to cosmic time $t$, and $Q = 6H^2$. Especially, for $f(Q) = -Q$ we retrieve the standard general relativity Friedmann's equations [28], as expected since as we have mentioned above, this specific choice for the functional form of the function $f(Q)$ is the STEGR limit of the theory. From Eqs. (15) and (16), we get the expressions of energy density $\rho$, pressure $p$ and equation of state (EoS) parameter $\omega$ respectively as

$$\rho = \frac{f}{2} - 6H^2 f_Q, \tag{17}$$

$$p = \left(\dot{H} + \frac{\dot{f_Q}}{f_Q}H\right)(2f_Q) - (\frac{f}{2} - 6H^2 f_Q), \tag{18}$$

$$\omega = \frac{p}{\rho} = -1 + \frac{\left(\dot{H} + \frac{\dot{f_Q}}{f_Q}H\right)(2f_Q)}{\left(\frac{f}{2} - 6H^2 f_Q\right)}. \tag{19}$$

Now, the trace of the field equations leads to the following conservation equation

$$\dot{\rho} + 3H(1+\omega)\rho = 0. \tag{20}$$

### III. COSMOLOGICAL $f(Q)$ MODEL

The system of field equations mentioned above has only two independent equations with four unknowns i.e. $\rho$, $p$, $f$, $H$. To solve the system totally and to examine the temporal evolution of energy density $\rho$, pressure $p$, and EoS parameter $\omega$, also necessity two constraint equations (supplementary conditions). In the literature, there are various arguments to select these equations. The procedure is well-known as the model-independent way technique to study DE cosmological models that generally consider a parameterization of any kinematic variables such as Hubble parameter and deceleration parameter, and provide the required additional equation. Following the same approach above, we take a parameterization of the deceleration parameter as [29]

$$q(z) = -\frac{\gamma + (z+1)^\gamma + 1}{(z+1)^\gamma + 1}, \tag{21}$$

where $\gamma$ is the model constant parameter. The motivation for choosing this form of deceleration parameter is that it is time-dependent and produces a transition from the early decelerating expansion phase to the current accelerating expansion phase for a certain range of negative values of $\gamma$ i.e. $-2 < \gamma < -1$. If $\gamma = -2$, this corresponds to a decelerating Universe in the present (i.e. at $z = 0$), for $\gamma < -2$, it corresponds to an accelerating Universe in the future (i.e. at $z < 0$), and in the case of $\gamma \geq -1$, this corresponds to an externally accelerating Universe.

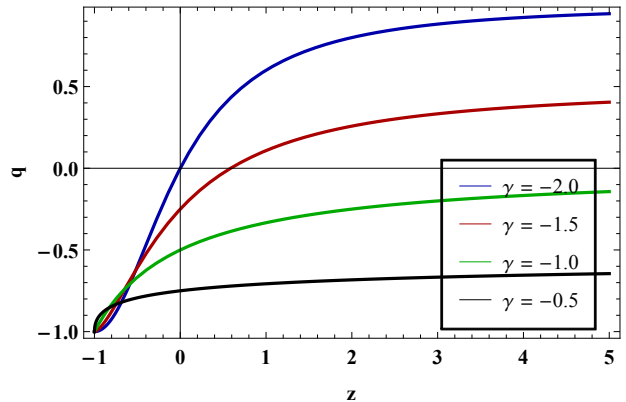

FIG. 1. *The plot of deceleration parameter versus z.*

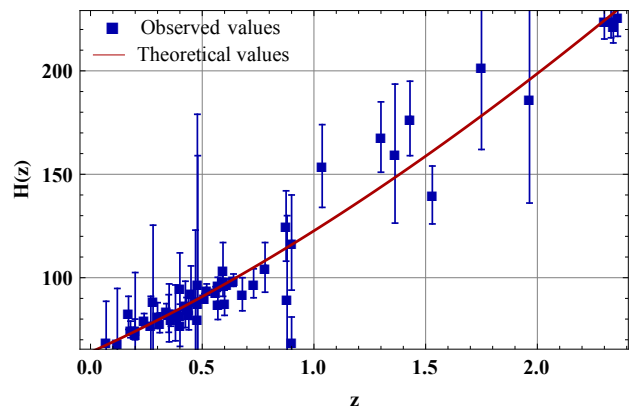

FIG. 2. *The plot of 57 points of H(z) datasets.*

The Hubble parameter determines the rate of expansion and acceleration or deceleration of the Universe. By integration Eq. (21) we find its expression in terms of redshift $z$ in the form

$$H(z) = \eta\left[(z+1)^{-\gamma} + 1\right], \tag{22}$$

where $\eta$ is an arbitrary constant of integration. $\eta$ is conducted to be positive, which ensures the positivity of the Hubble parameter regardless of the value of the constant $\gamma$. The deceleration parameter as a function of Hubble parameter $H(z)$ is given by $q = -1 - \frac{\dot{H}}{H^2}$. If $q > 0$, the Universe is in a decelerated phase, else $q < 0$ conform to an accelerated phase. Fig. 1 shows the behavior of $q(z)$ in terms of redshift $z$, explaining the evolution from past to present for certain values of $\gamma$. In the case of $\gamma = -1.5$, our model transitions from a positive value in the past, i.e. an early deceleration, to a negative value in the present, indicating a current acceleration, and finally tends to $-1$. Moreover, the current value of $q$ obtained corresponds to the observation data.

To constrain the model parameters $\gamma$ and $\eta$ and to compare our results with observations. Here, in this investigation, we have used the Hubble datasets with 57 data points in the redshift range $0.07 \leq z \leq 2.42$ (see Tab. I). In this set of 57 Hubble data points, 31 points were measured via the method of differential age (DA) and the remaining 26 points through BAO (Baryonic Acoustic Oscillations) and other methods [30–32]. To find the best fit value of the model parameters of our obtained models, we have used the technique $R^2$-test defined by following the statistical formula

$$R^2_{OHD} = 1 - \frac{\sum_{i=1}^{57}\left[(H_i)_{obs} - (H_i)_{th}\right]^2}{\sum_{i=1}^{57}\left[(H_i)_{obs} - (H_i)_{mean}\right]^2}, \quad (23)$$

where, $(H_i)_{obs}$ and $(H_i)_{th}$ are observed and predicted values of Hubble parameter respectively. The observed values are conducted from Tab. I and theoretical values are considered from Eq. (22). Although the statistical fitting using the technique $R^2$-test is rather basic, it is adequate for the illustrative purposes of the paper. For maximum value of $R^2_{OHD}$, we obtain the best fit values of the model parameters $\gamma = -1.502$ and $\eta = 32.01$ with $R^2_{OHD} = 0.9334$ and root mean square error $RMSE = 11.44$ and their $R^2_{OHD}$ values only 6.66% far from the best one. The perfec case $R^2_{OHD} = 1$ occurs when the observed data and theoretical function $H(z)$ agree precisely. By replacing the values of $\gamma$ and $\eta$ into Eq. (22) and at $z = 0$ we find the current value of the Hubble parameter $H$ for our model as $H_0 = 64.02 \pm 11.44 Km/s/Mpc$. Fig. 2 shows the best fit curve of the Hubble parameter $H(z)$ versus redshift $z$ using 57 Hubble parameter measurements.

To add a second constraint on the field equations, Mandal et al. [18] discussed the energy conditions for $f(Q)$ gravity by assuming the function $f(Q)$ takes the form of a algebraic polynomial function and a logarithmic function of the non-metricity $Q$. Further, in another work, the scenario of the Universe accelerating in $f(Q)$ gravity was checked using the logarithmic function [33]. This investigation, motivated us to work with a logarithmic dependence of $f(Q)$ gravity with the model or free parameters $\alpha$ and $\beta$ i.e.

$$f(Q) = \alpha + \beta \log Q. \quad (24)$$

Using Eq. (24) in Eq. (15) and Eq. (16), the expressions of energy density $\rho$, pressure $p$ and EoS parameter $\omega$ are given respectively as

$$\rho(z) = \frac{\alpha}{2} + \frac{\beta}{2}\left[\log\left(6\eta^2\left[1 + \frac{1}{(1+z)^\gamma}\right]^2\right) - 2\right], \quad (25)$$

$$p(z) = -\frac{\alpha}{2} - \frac{\beta}{3}\left\{\frac{\gamma}{1+(1+z)^\gamma}\right\} - \frac{\beta}{2}\left[\log\left(6\eta^2\left[1 + \frac{1}{(1+z)^\gamma}\right]^2\right) - 2\right], \quad (26)$$

$$\omega(z) = -1 - \frac{\beta}{3} \times \left\{\frac{\gamma}{1+(1+z)^\gamma}\right\} \times \left\{\frac{\alpha}{2} + \frac{\beta}{2}\left[\log\left(6\eta^2\left[1 + \frac{1}{(1+z)^\gamma}\right]^2\right) - 2\right]\right\}^{-1}. \quad (27)$$

The behavior of energy density $\rho(z)$ and pressure $p(z)$ of our model is shown in Figs. 3 and 4. It is observed that energy density is a increasing function of redshift $z$, and the pressure takes negative values during the cosmic evolution. The negative pressure is due to the logarithmic form considered, which indicates the expanding accelerated phase of the Universe. Fig. 5 shows the evolution of the EoS parameter $\omega(z)$ for the logarithmic model versus redshift $z$. It can be observed that $\omega(z)$ of our model varies with the quintessence zone $(-1 < \omega(z) < -1/3)$ throughout its evolution of all values of $\gamma$. In addition, we can see that the EoS parameter approaches $\Lambda CDM$ model $(\omega(z) = -1)$ in the future i.e. $z \to -1$, which leads to the same behavior in these references [34]. Further, the values of EoS parameter are consistent with the observational data from several experiments [3].

The energy conditions are a way to validate dark energy models. In general relativity, energy conditions play an important role in proving theories about the existence of the space-time singularity and black holes [35]. In the context of dark energy, these conditions are used to verify the accelerating expansion of the Universe. The energy conditions in symmetric teleparallel gravity are discussed in [18]. In this work, we discuss some of the most common energy conditions for this model. Namely, the weak energy conditions (WEC), the dominant energy conditions (DEC), the null energy conditions (NEC) and the strong energy conditions (SEC), we yield the constraints as

- WEC: if $\rho \geq 0, p + \rho \geq 0$;
- DEC: if $\rho \geq 0, |p| \leq \rho$;
- NEC: if $p + \rho \geq 0$;
- SEC: if $\rho + 3p \geq 0$.

For our model these energy conditions are obtained as

$$\text{DEC: } \rho - p = \alpha + \frac{\beta}{3}\left\{\frac{\gamma}{1+(1+z)^\gamma}\right\} + \beta\left[\log\left(6\eta^2\left[1+\frac{1}{(1+z)^\gamma}\right]^2\right) - 2\right] \geq 0, \tag{28}$$

$$\text{NEC: } \rho + p = -\frac{\beta}{3}\left\{\frac{\gamma}{1+(1+z)^\gamma}\right\} \geq 0, \tag{29}$$

$$\text{SEC: } \rho + 3p = -\alpha - \beta\left\{\frac{\gamma}{1+(1+z)^\gamma}\right\} - \beta\left[\log\left(6\eta^2\left[1+\frac{1}{(1+z)^\gamma}\right]^2\right) - 2\right] \geq 0. \tag{30}$$

Figs. 6, 7, and 8 represent the energy conditions for the logarithmic form as a function of redshift. In the present study, it is observed that NEC, WEC and DEC are well satisfied throughout cosmic evolution. However, the SEC is violated for our model. The violation of the SEC is due to the accelerating expansion of the Universe.

### IV. PHYSICAL PROPERTIES OF THE MODEL

#### A. Jerk parameter

In order to study the physical properties of the model, we start from an important parameter to describe the dynamics of the Universe, which is called the jerk parameter. The advantage of this parameter is that it makes possible to describe models close to the $\Lambda CDM$ model. In the literature [37], the transition from acceleration to deceleration of the Universe occurs for different models with a positive value of the jerk parameter and a negative value of the deceleration parameter. In the flat $\Lambda CDM$ model, the value of the jerk parameter $j = 1$. The jerk parameter is defined as [38]

$$j = \frac{\dddot{a}}{aH^3}. \tag{31}$$

where, $a(t) = \left(\exp(-\gamma\eta t) - 1\right)^{\frac{-1}{\gamma}}$ is the scale factor of the model and its relationship to the redshift is given by the form $a(t) = \frac{1}{1+z}$. Thus, the expression of jerk parameter for our model is

$$j(z) = 1 + \frac{\gamma\left(\frac{1}{1+z}\right)^\gamma \left\{3 + \gamma + \left(\frac{1}{1+z}\right)^\gamma (3+2\gamma)\right\}}{\left\{1 + \left(\frac{1}{1+z}\right)^\gamma\right\}^2}. \tag{32}$$

Fig. 9 shows that the cosmic jerk parameter is positive throughout cosmic evolution and tends to 1 in later times (i.e. at $z \to -1$), which indicates the consistency of our model with the $\Lambda CDM$ model in the future.

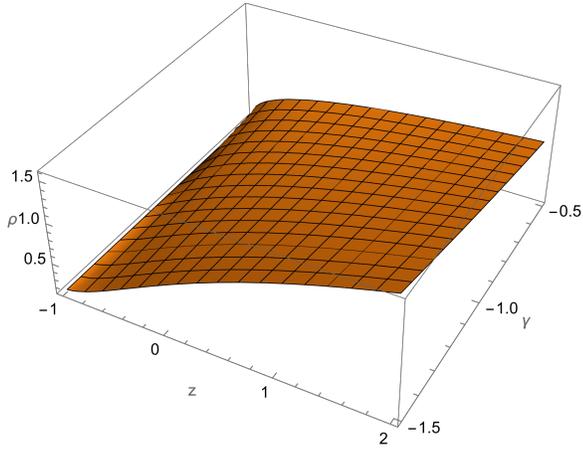

FIG. 3. *The plot of energy density versus z and $\gamma$ with $\eta = 32$, $\alpha = -5$ and $\beta = 0.788$.*

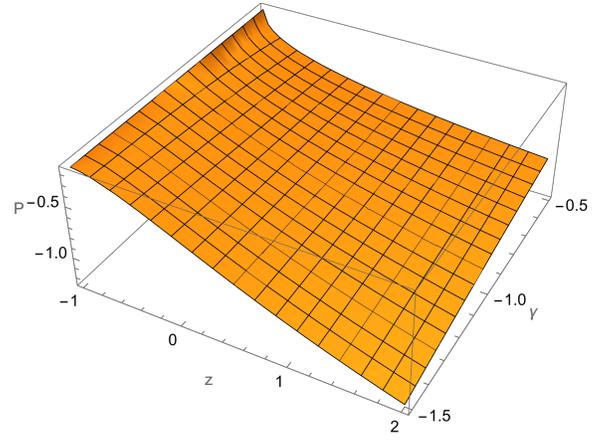

FIG. 4. *The plot of pressure versus z and $\gamma$ with $\eta = 32$, $\alpha = -5$ and $\beta = 0.788$.*

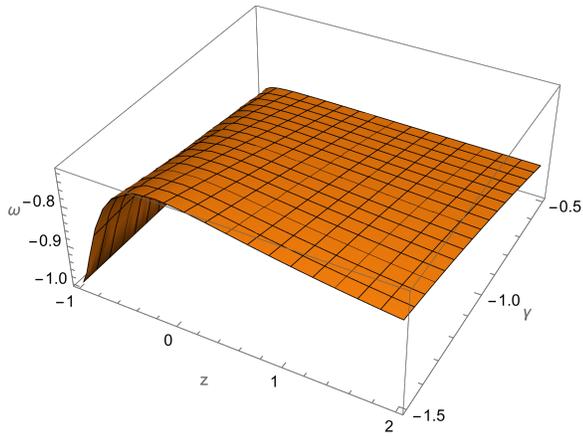

FIG. 5. *The plot of EoS parameter versus z and $\gamma$ with $\eta = 32$, $\alpha = -5$ and $\beta = 0.788$.*

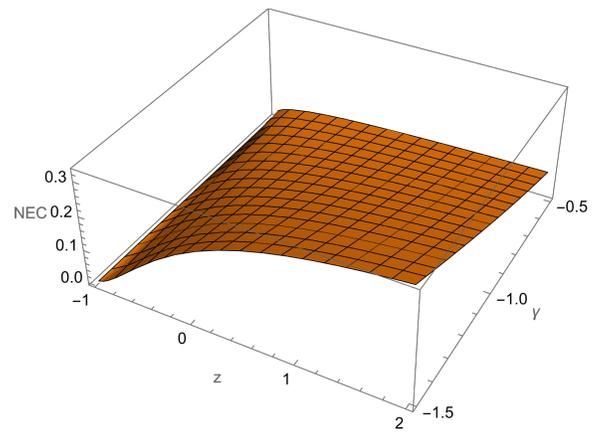

FIG. 6. *The plot of NEC versus z and $\gamma$ with $\eta = 32$, $\alpha = -5$ and $\beta = 0.788$.*

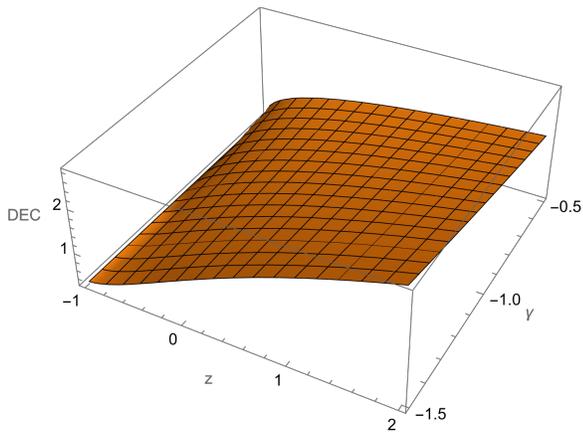

FIG. 7. *The plot of DEC versus z and $\gamma$ with $\eta = 32$, $\alpha = -5$ and $\beta = 0.788$.*

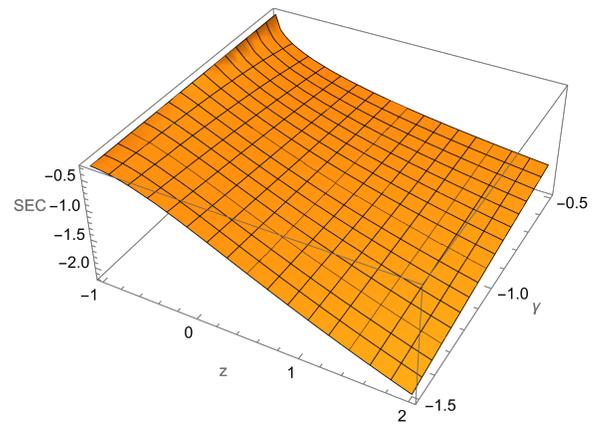

FIG. 8. *The plot of SEC versus z and $\gamma$ with $\eta = 32$, $\alpha = -5$ and $\beta = 0.788$.*

## B. Statefinder parameters

The statefinder paramaters $\{r,s\}$ is defined as

$$r = \frac{\dddot{a}}{aH^3}, \quad s = \frac{r-1}{3\left(q-\frac{1}{2}\right)}, \left(q \neq \frac{1}{2}\right). \quad (33)$$

These parameters were introduced by Sahni et al. [39] to distinguish dark energy models like the $\Lambda$CDM, SCDM (Standard Cold Dark Matter), HDE (Holographic Dark Energy), CG (Chaplygin Gas) and Quintessence. Here, $r$ is the same as the jerk parameter in the previous section. Different values for the pair $\{r,s\}$ represent different dark energy models, we mention in particular

- For $\Lambda$CDM$\rightarrow (r=1, s=0)$,
- For SCDM$\rightarrow (r=1, s=1)$,
- For HDE$\rightarrow \left(r=1, s=\frac{2}{3}\right)$,
- For CG$\rightarrow (r>1, s<0)$,
- For Quintessence$\rightarrow (r<1, s>0)$.

For the presented model, the statefinder parameters in terms of redshift $z$ are derived as

$$r(z) = 1 + \frac{\gamma\left(\frac{1}{1+z}\right)^{\gamma}\left\{3+\gamma+\left(\frac{1}{1+z}\right)^{\gamma}(3+2\gamma)\right\}}{\left\{1+\left(\frac{1}{1+z}\right)^{\gamma}\right\}^2}, \quad (34)$$

$$s(z) = \frac{\gamma}{3}\left\{-2 + \frac{1}{1+\left(\frac{1}{1+z}\right)^{\gamma}} + \frac{3}{3+\left(\frac{1}{1+z}\right)^{\gamma}(3+2\gamma)}\right\}. \quad (35)$$

This procedure for differentiating between dark energy models is model-independent because it does not require any knowledge around the cosmological model. Here, we have plotted the $\{r,s\}$ trajectory in the $s-r$ plane represented in Fig. 10 with $\gamma = -1.5$. From this figure, it is clear that the model evolve from the Quintessence region $(r<1, s>0)$ at present, and finally tends to the $\Lambda CDM$ model $(r=1, s=0)$. Hence, the created model behaves as a quintessence dark energy model at present.

## V. DISCUSSION AND CONCLUSION

In this work, we have discussed the present and late time cosmological homogeneous and isotropic flat FLRW Universe take on modified $f(Q)$ gravity model with the functional form as $f(Q) = \alpha + \beta \log(Q)$ where $\alpha$ and $\beta$ are the model constant parameters. In this analysis we constrain the model constant parameters using observational datasets of the updated 57 points of Hubble datasets in which 31 points measured via the method of differential age (DA) and remaining 26 points through BAO and other methods, we find the deceleration parameter to be negative and consistent with the present scenario of an accelerating Universe. For the parametrization of the deceleration parameter we use an approach as $q(z) = -\frac{\gamma+(z+1)^{\gamma}+1}{(z+1)^{\gamma}+1}$ where $\gamma$ be any constant. This form of deceleration parameter is time-dependent and produces a transition from the early decelerating expansion phase to the current accelerating expansion phase for a certain range of values of $\gamma$. For the said form of deceleration parameter we found the Hubble parameter which determines the rate of expansion of the Universe. Also, it is observed that for the choice of $\gamma < -1.5$, the model transits from a positive value in the past i.e. an early deceleration, to a negative value in the present, indicating a current acceleration and finally tends to $-1$ moreover the current value of deceleration parameter corresponds to the observational data. The pressure in the model takes negative value during the cosmic evolution which is due to the considered logarithmic form of $f$, represents the expanding accelerated phase of the Universe. The EoS parameter in the model acquired $(\omega(z) = -1)$ in the future from the quintessence zone throughout its evolution which characterises that in the future the model approaches $\Lambda CDM$ model after quintessence zone and the values of $\omega(z)$ parameter are consistent with the observational data from several experiments.

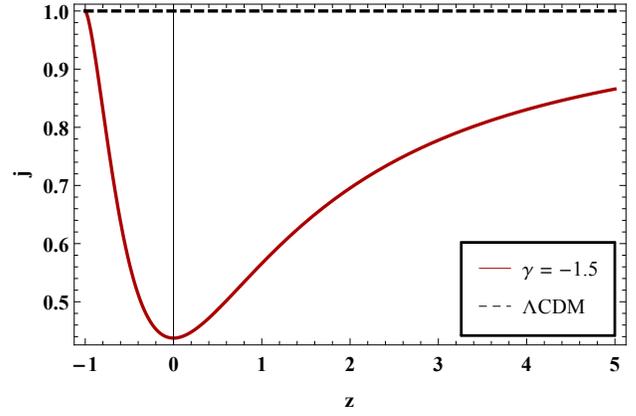

FIG. 9. *The plot of jerk parameter j(z) versus redshift z*

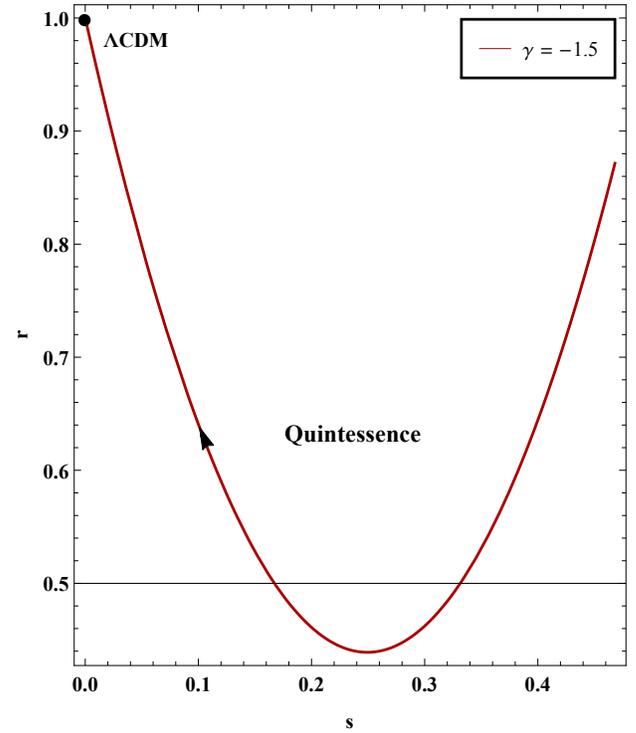

FIG. 10. *The plot of (s, r) trajectories*

| $z$ | $H(z)$ | $\sigma_H$ | $z$ | $H(z)$ | $\sigma_H$ |
|---|---|---|---|---|---|
| 0.070 | 69 | 19.6 | 0.4783 | 80 | 99 |
| 0.90 | 69 | 12 | 0.480 | 97 | 62 |
| 0.120 | 68.6 | 26.2 | 0.593 | 104 | 13 |
| 0.170 | 83 | 8 | 0.6797 | 92 | 8 |
| 0.1791 | 75 | 4 | 0.7812 | 105 | 12 |
| 0.1993 | 75 | 5 | 0.8754 | 125 | 17 |
| 0.200 | 72.9 | 29.6 | 0.880 | 90 | 40 |
| 0.270 | 77 | 14 | 0.900 | 117 | 23 |
| 0.280 | 88.8 | 36.6 | 1.037 | 154 | 20 |
| 0.3519 | 83 | 14 | 1.300 | 168 | 17 |
| 0.3802 | 83 | 13.5 | 1.363 | 160 | 33.6 |
| 0.400 | 95 | 17 | 1.430 | 177 | 18 |
| 0.4004 | 77 | 10.2 | 1.530 | 140 | 14 |
| 0.4247 | 87.1 | 11.2 | 1.750 | 202 | 40 |
| 0.4497 | 92.8 | 12.9 | 1.965 | 186.5 | 50.4 |
| 0.470 | 89 | 34 | | | |
| $z$ | $H(z)$ | $\sigma_H$ | $z$ | $H(z)$ | $\sigma_H$ |
| 0.24 | 79.69 | 2.99 | 0.52 | 94.35 | 2.64 |
| 0.30 | 81.7 | 6.22 | 0.56 | 93.34 | 2.3 |
| 0.31 | 78.18 | 4.74 | 0.57 | 87.6 | 7.8 |
| 0.34 | 83.8 | 3.66 | 0.57 | 96.8 | 3.4 |
| 0.35 | 82.7 | 9.1 | 0.59 | 98.48 | 3.18 |
| 0.36 | 79.94 | 3.38 | 0.60 | 87.9 | 6.1 |
| 0.38 | 81.5 | 1.9 | 0.61 | 97.3 | 2.1 |
| 0.40 | 82.04 | 2.03 | 0.64 | 98.82 | 2.98 |
| 0.43 | 86.45 | 3.97 | 0.73 | 97.3 | 7.0 |
| 0.44 | 82.6 | 7.8 | 2.30 | 224 | 8.6 |
| 0.44 | 84.81 | 1.83 | 2.33 | 224 | 8 |
| 0.48 | 87.90 | 2.03 | 2.34 | 222 | 8.5 |
| 0.51 | 90.4 | 1.9 | 2.36 | 226 | 9.3 |

TABLE I. 57 points of $H(z)$ data: 31 (DA) + 26 (BAO+other) [30].

For the obtained Hubble parameter we used the Hubble datasets with 57 data points for the redshift range $0.07 \leq z \leq 2.42$ presented in Tab. I using the technique $R^2$-test to find the best fit value of the model parameters and observed the best fit values of the model parameters $\gamma = -1.502$ and $\eta = 32.01$ with $R^2_{OHD} = 0.9334$ and root mean square error 11.44 and their $R^2_{OHD}$ values only 6.66% far from the best one.

For the validation of model one we use the tool as energy conditions, this conditions are used to verify the accelerating expansion of the Universe. We discuss some of the most common energy conditions like WEC, DEC and SEC for this model and observed that WEC and DEC both are well satisfied throughout cosmic evolution, however the SEC is violated. The violation of the SEC validate an accelerating expansion of the Universe. Also, we discussed the Jerk and statefinder parameters and observed that the Jerk parameter is $j = 1$ point toward $\Lambda CDM$ model and the state finder parameters are $(r, s) \equiv (1, 0)$ which also point towards $\Lambda CDM$ model in future.

Finally, it is interesting to note that the negative pressure in our cosmological model is incompatible with ordinary matter and demonstrates the need for exotic matter to explain the accelerating expansion of the current Universe. In addition, this can be seen by the SEC ($\rho + 3p \geq 0$) conditions that appears to be violated. Thus, we conclude that the modifications to gravity do not eliminate the need for exotic matter.

### ACKNOWLEDGMENTS

We are very much grateful to the editor and anonymous referee for illuminating suggestions that have significantly improved our article.